# Some Pattern Recognition Challenges in Data-Intensive Astronomy★

S. G. Djorgovski, C. Donalek, A. Mahabal, R. Williams, A.J. Drake, M.J. Graham, E. Glikman
*Astronomy Dep. and Ctr. for Advanced Computing Research, Caltech, Pasadena, CA 91125, USA*
*[george, donalek, aam, rwilliams, ajd, mjg, eilatg] @ astro.caltech.edu*

**Abstract**

*We review some of the recent developments and challenges posed by the data analysis in modern digital sky surveys, which are representative of the information-rich astronomy in the context of Virtual Observatory. Illustrative examples include the problems of an automated star-galaxy classification in complex and heterogeneous panoramic imaging data sets, and an automated, iterative, dynamical classification of transient events detected in synoptic sky surveys. These problems offer good opportunities for productive collaborations between astronomers and applied computer scientists and statisticians, and are representative of the kind of challenges now present in all data-intensive fields. We discuss briefly some emergent types of scalable scientific data analysis systems with a broad applicability.*

## 1. Challenges and Opportunities of Data-Rich Astronomy and Other Sciences

Like nearly every other field of science, astronomy is now facing an exponential growth in the volume, complexity and even quality of data, both from actual measurements (e.g., massive digital sky surveys) and from numerical simulations of processes and phenomena which cannot be addressed in a simple analytical fashion (e.g., structure formation in the universe, supernova explosions, etc.) [1,2]. This exponential growth is driven by the progress in information technology (IT), and consequently we see doubling of the information volume in astronomy every 12 – 18 months, with about 1 PB currently archived, and the data growth rate of ~ 2 TB/day for the astronomy worldwide. Data sets measured in tens of TB are now becoming common, and multi-PB data sets are on the horizon, in particular in the form of large synoptic sky surveys.

This explosive growth of information has a great enabling power – provided that the richness of the newly available data can be managed, explored and analysed in an effective manner. This is a very non-trivial task. There is also a great commonality of data handling and understanding challenges across all scientific disciplines, as well as other fields: the modern commerce, finance, security, etc. [3]

The astronomical community has responded to these challenges with the concept of a Virtual Observatory (VO): a geographically and institutionally distributed, web-based research environment for astronomy with massive and complex data sets, which unifies data archives and other information infrastructure, and computational and data analysis tools for their exploration and analysis [4,5,6,7]. A number of national VO's as well as a vibrant international alliance of them are now active [8,9].

Similar types of virtual scientific organizations have been created in many other fields, and more are appearing constantly; they are discipline-based, rather than institution- or agency-based. All are parts of the new cyberinfrastructure of science [10], and there is probably some avoidable duplication of efforts.

In the VO community, there has been an excellent progress in the matters of data management: archives, standards, protocols, interoperability, etc. However, there has been relatively little progress in the development of highly scalable data exploration and analysis tools needed to generate the scientific returns from these large and expensively obtained data sets.

While there are many off-the-shelf data mining tools and systems available, few if any of them can really scale effectively to TB and PB size data sets. High statistical dimensionality and complexity present even larger technical challenges than the data volumes alone. The lack of such tools, and the resulting scarcity of scientific results, has delayed a broader community buy-in into these developments. This is perhaps the focal problem of eScience (aka Cyberscience) today.

Here we review some examples of data analysis challenges faced by astronomers, using mainly the new Palomar-Quest digital synoptic sky survey [11] as a test case. These problems are neither trivial nor hopelessly difficult, and they provide a great opportunity for collaborations between astronomers and applied computer scientists and IT professionals. We hope to stimulate such collaborations.

## 2. Star-Galaxy Image Classification: The Next Generation

A classical problem in the analysis of astronomical panoramic imagery – most notably large sky surveys, which are now the largest source of astronomical data by volume – is morphological classification of detected sources. At the most basic level, this is the separation of sources into spatially unresolved ones ("stars", but physically also including quasars or possibly other types of objects whose apparent angular size is much smaller than the effective angular resolution of the survey), and spatially resolved ones ("galaxies", but possibly other kinds of nebulae). The accuracy and completeness of the morphological source classification is often the limiting factor in the scientific applications of such data, more stringent than the detection (e.g., flux) limits. The characteristic resolution of an astronomical image is given through a combination of the instrumental resolution (usually given by the optics) and atmospheric turbulence for the ground-based optical and IR surveys, resulting in a net point-spread function (PSF) or "beam", to which astronomers often refer to as the "seeing". Image sampling is generally matched to the typical PSF, with the Nyquist sampling or better.

The problem can be stated as: is any given source well described to within the measurement errors by the PSF (a "star"), or is it significantly more extended (a "galaxy")? And what is the probability of belonging to either class? A more sophisticated classification also allows for probable image artifacts, and for the well resolved sources a secondary classification (e.g., galaxy types) may be applied.

In the case of homogeneous imagery, this problem has been solved fairly well over a decade ago [e.g., 12,13,14,15, and refs. therein]. Typical approaches include simple dividers in some parameter space, or application of supervised classification or machine learning (ML) tools such as the artificial neural nets (ANN) or decision trees (DT). For example, in the PQ survey, we use a Multilayer Perceptron in a Bayesian framework, with a softmax activation function and 2 output nodes, one for stars and one for galaxies. Data sets with a superior angular resolution and depth, where accurate source classifications can be obtained simply by a visual inspection and/or spectroscopic confirmation, are used as training and testing data sets.

Typically each individual image (e.g., one of the many CCD frames or photographic plate scans comprising the survey) is treated independently from others, even though some useful information is present in the neighboring images, which generally have very similar properties. Also, in a survey, homogeneity of morphological classification is important, and has to be achieved by normalizing the object attributes fed into the classifiers.

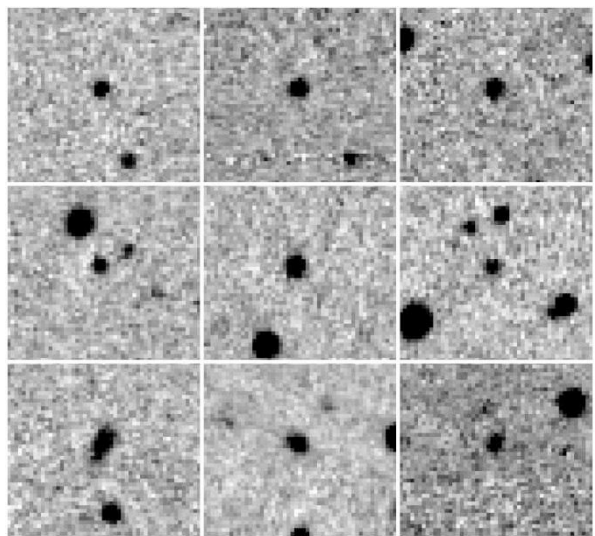

Figure 1. Examples of sources from the Palomar-Quest survey, classified using ANN techniques Top row: sources classified as stars with a probability $p^* > 90\%$; bottom row: sources classified as galaxies, with $p^* < 10\%$; middle row: intermediate-classification sources with $p^* \approx 50\%$.

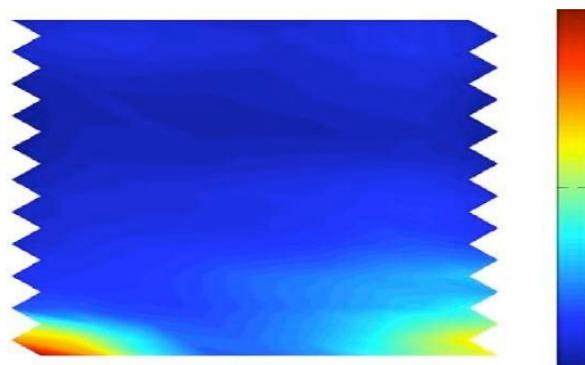

Figure 2. ANN classification and performance can be analysed using Kohonen's self-organizing maps (SOM). This can yield information about the consistency of classifications, the relative importance of various input parameters, etc. We use such methods for the optimization of ANN classifiers used in the PQ survey. From Donalek et al., in prep.

There have been also some initial exploratory applications of unsupervised classification methods

[13,16,17,18,19]. They are potentially useful for more elaborate exploration of data, but if we can decide a priori on the number of classes (in this case, two: sources are either unresolved or not), then a supervised classifier may be better.

We note that the same types of classification techniques are also used for more detailed explorations of large digital sky surveys and other astronomical data sets, especially in searches for outliers in some parameter space, which are often some astrophysically interesting type of objects (e.g., distant quasars) [20,21,22,23,24,37,40]. Resolved sources ("galaxies") in principle contain more morphological information, since all unresolved ones ("stars") by definition look alike. Thus, for the sufficiently well resolved objects one can apply a secondary morphological classification, e.g., determining the galaxy types.

With the advent of modern, multi-bandpass or synoptic (multi-epoch) sky surveys, and federation of multiple sky surveys and other data sets in the VO context, often with a variety of resolutions, PSFs, data quality, etc., the problem of star-galaxy classification has come back at a much more complex level.

Essentially, the same astronomical source is being imaged many times in different conditions, different filters, etc., and it can get many independent classifications, which need not be mutually consistent. Yet, any given source is intrinsically either a "star" (including quasars) or a "galaxy" (or some other nebulosity). Given the available abundance of heterogeneous data, what is the optimal joint classification for any given source?[1] The choice may be also scientific application dependent, not just data dependent.

One approach to this problem is to associate a degree of reliability to each of the independently derived classes for a given source, giving them appropriate statistical weights (e.g., depending on the S/N or the PSF), and then performing some optimized meta-classification. Alternatively, one could try to perform a classification process using all measured parameters from all imaging passes at once. As of this writing, it is not known how to do this well, and what are the advantages and disadvantages of each approach.

---

[1] There is a similar, but perhaps easier problem of optimized source detection given multiple images in different filters, from different instruments, at different times, etc. A source may be detected with varying degrees of statistical significance in some of them, but not in others; this could be due to the variations in the data quality and/or the intrinsic variability or the spectral energy distribution; in many cases non-detections also provide useful information. One has to evaluate a joint significance for multiple detections. For an interesting approach to this problem, see [25].

Another problem can be termed *context-based image classification*. In addition to the information present in an image in which sources are being classified, there is also some useful external, a priori information which could and should be used. For example, we expect that the relative fractions of stars and galaxies will change continuously across the sky and as a function of the flux; there should be no discrete jumps in the ratio of stars to galaxies across the edges of adjacent images. Another constraint may come from domain knowledge, e.g., that the relative fraction of stars would be nearly 100% in the direction of the Galactic plane, but much lower at the Galactic poles. We need a way to incorporate such external constraints without introduction of biases which would affect the physical interpretation of the data. One possibility is to design and implement a suitable cost function into the classification algorithm.

## 3. Dynamical, Real-Time Classification of Astronomical Transient Events in Synoptic Sky Surveys

The scientific measurement and discovery process and method traditionally follows the pattern of theory followed by experiment, analysis of results, and then follow-up experiments, often on time scales from days to decades after the original measurements, feeding back to a new theoretical understanding. But what about phenomena where a rapid change occurs on time scales shorter than what it takes to set up the new round of measurements? Thus a need for dynamical, real-time scientific measurement systems, consisting of discovery instruments or sensors, a real-time computational analysis and decision engine, and optimized follow-up instruments which can be deployed selectively in (or in near) real-time, where measurements feed back into the analysis immediately.

In astronomy, examples of rapidly changing phenomena or transient events include supernovae, gamma-ray bursts, gravitational microlensing events, planetary occultations, stellar flares, accretion flares from supermassive black holes, rapidly moving potential planetary hazard asteroids, and in the future gravitational wave bursts, etc. The time domain is rapidly becoming one of the most exciting new research frontiers in astronomy. The sky is no longer seen as a slowly and orderly changing; there are important physical phenomena occurring on scales as short as seconds, whose rapid and appropriate follow-up promises to broaden substantially our understanding of the physical universe, and perhaps lead to a discovery of previously unknown phenomena.

A number of astronomical surveys and time-domain experiments are already operating [see, e.g., 26,27,28, 29,30,35,36], and much more ambitious enterprises are being planned [31,32], with multi-TB data streams, which will be yielding hundreds or thousands of transient events per night, implying a need for automated, robust processing and follow-up. There is a growing number of autonomous robotic telescopes geared to discovery and follow-up of transient events. Yet, most systems rely on a delayed human judgment in decision making and follow-up of events.

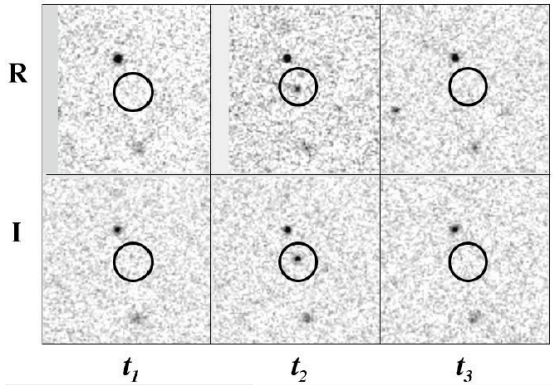

**Figure 3.** Examples of transient events from the Palomar-Quest survey. A source appears at the second epoch ($t_1$, middle column) in both filters ($R$, top, and $I$, bottom), whereas it is not present at two other epochs ($t_2$ and $t_3$), separated by some weeks. Many such events are discovered archivally, and their physical nature is still unknown. Real-time detection and follow-up are necessary to gain more insights.

The scientific driver here is to classify the transient events in terms of their physical nature, on the basis of the observed patterns of their variability, spectrum, and other attributes. This would then trigger or drive follow-up observations and scientific interpretation.

We are developing a system which would address the need for a rapid and automated discovery and follow-up of transient events, called VOEventNet (PI: R. Williams) [38]. The system uses the emerging VOEvent communication protocols for information exchange between the principal event discovery and follow-up engine, fed primarily by the Palomar-Quest (PQ) survey data stream in the real time, and several partnering robotic observatories, which may react to the event alerts based on their own selection criteria.

The system uses the primary event measurement data from PQ, archival data from previous passes on the same part of the sky, archives of known asteroids and variable stars, quasars, etc., and a broad array of VO-connected archives and surveys to generate the initial dataset on which to base the preliminary event classification used by the potential follow-up facilities.

An essential feature of the system is the feedback loop which incorporates new follow-up data from affiliated robotic telescopes (and indeed from any other external source), which are folded into a dynamically evolving classification for each event. All of the pertinent data are posted in the real time on an open website, and alerts are distributed to a subscriber list.

Consumers of transient events are usually interested only in a particular type, e.g., supernovae usable as cosmological standard candles, microlensing events, etc. Thus the desired output is to evaluate a probability of any given event as belonging to any of the possible known classes. The most interesting outcome may be the events which do not fit any of the known patterns – possible examples of new types of astronomical objects or phenomena.

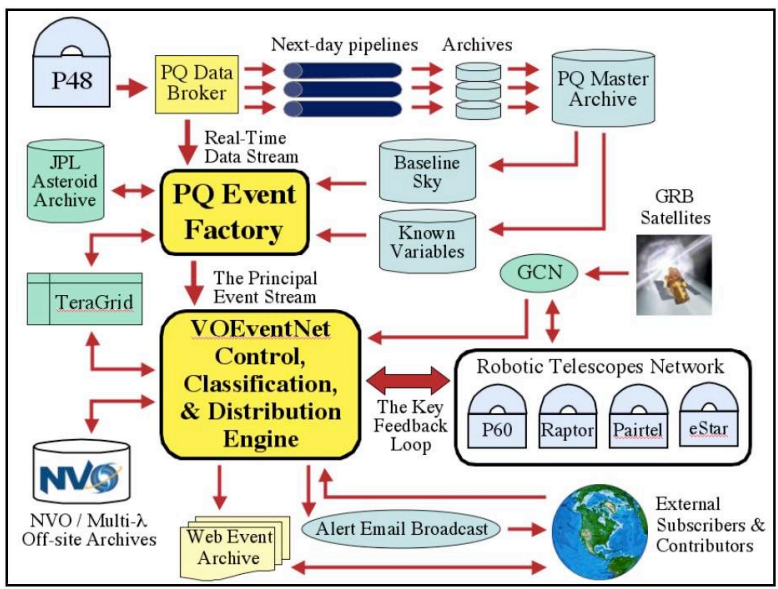

**Figure 4.** A schematic outline of the VOEventNet system for an automated, dynamical, real-time discovery, classification, distribution, and follow-up of astronomical transient events. The system is fed by areal-time data stream from the PQ survey. Transients are discovered by comparing the current data with a baseline sky, and classified ion the basis of new measurements and archival data sets.

The goal here is to associate classification probabilities for any given event as belonging to a variety of known classes of variable astrophysical objects (e.g., quasars, stellar explosions, variable stars, etc.) and to update such classifications as more data come in, until a scientifically justified convergence.

The process has to be fully automated, robust, and reliable; it has to operate from sparse and heterogeneous data; it has to maintain a high modern scientific data analysis systems, which could benefit single or multiple disciplines or projects.

The universal applicability or functionality of such systems derives from the commonality of computational or data related challenges: everyone needs properly archived, annotated and indexed data sets, data discovery and access tools, data fusion mechanisms, and a broad variety of data mining and visualization methods.

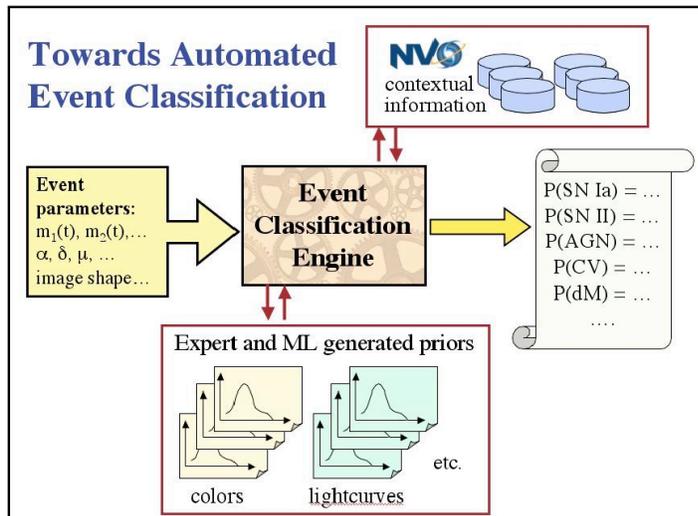

Figure 5. A schematic outline of the automated event classification engine. The input consists of the generally sparse discovery data, including brightness in various filters, possibly the rate of change, position, possible motion on the sky, etc., and measurements from external, multi-wavelength archives corresponding to this spatial location, if available; and a library of priors giving probabilities for observing these particular parameters if the event was belonging to a class X. The output is an evolving set of probabilities of belonging to various classes of interest.

completeness (not miss any interesting events) yet a low false alarm rate; and it has to learn from the past experience. These are very challenging requirements, and a pattern recognition problem par excellance!

One approach is to generate (and update) a library of prior distributions of the type "if this was a supernova of the type Ia, the probability of changing brightness by this much in this filter over this time interval is such and such", and to do it for a broad variety of known variable astrophysical phenomena (and this knowledge is bound to be incomplete!). Then a synthetic probability of an event belonging to any given (known) class would be evaluated from all of such pieces of information available, perhaps in some Bayesian fashion. This poses many conceptual and technical challenges, and the work is now just starting.

## 4. A New Generation of Scientific Data Analysis and Exploration Systems

While the VO and equivalent organizations serve the broad needs of their scientific communities, and individual computationally-intensive or data-intensive experiments or projects develop customized data processing systems for their own specific needs, there may be a growing need for an intermediate level of

One such system for exploration of Petascale scientific data sets is being currently designed at Caltech Center for Advanced Computing Research (CACR); M. Stalzer is the PI, and several of the present authors are involved, along with collaborators from other disciplines and computer science. The goal is to develop a standardized data architecture in which large data sets from a broad variety of sources (including astronomy, biology, geophysics, etc.) can be ingested and prepared for data mining; and a variety of highly scalable data mining and hyperdimensional visualization tools which could then be applied to them for knowledge extraction.

Another approach is to harness the power of distributed computational resources through discipline-specific *Grid Science Gateways* [33]. The goal here is to make the power of the Grid computing more accessible to domain experts who may not be familiar with the technical arcana which pervades the current Grid activity, and to do it through some transparent, graduated security approach.

An example of such an approach is the GRIST project [34, 39] at Caltech and JPL (R. Williams is the PI). The goal is to separate data flow from process control (workflow), and provide an user with a friendly and flexible interface with which to connect available data sources and streams, computational tools, and data mining and visualization algorithms. Many of these are deployed as web services in a Grid environment.

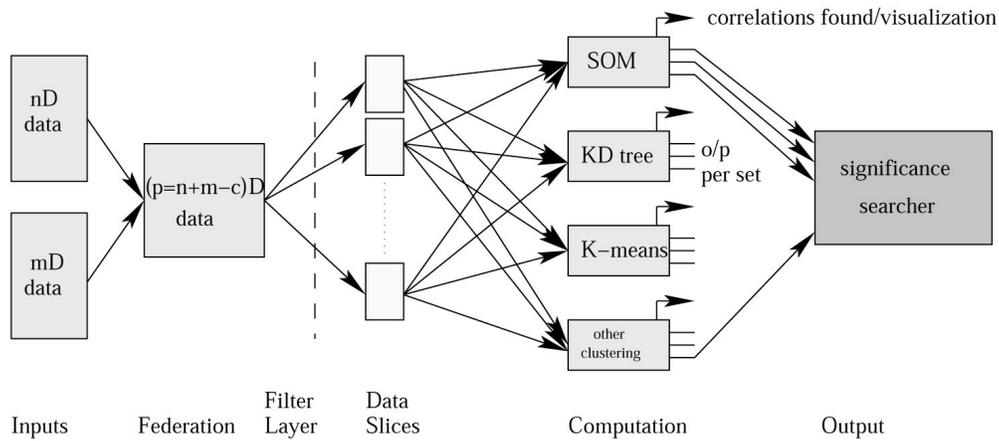

Figure 6. A schematic example of a hypothetical GRIST-based data analysis application, in which several statistical dimensionality reduction techniques and clustering algorithms are used to partition fusion of complex data sets into smaller, significant units/clusters, and are fed into subsequent analysis.

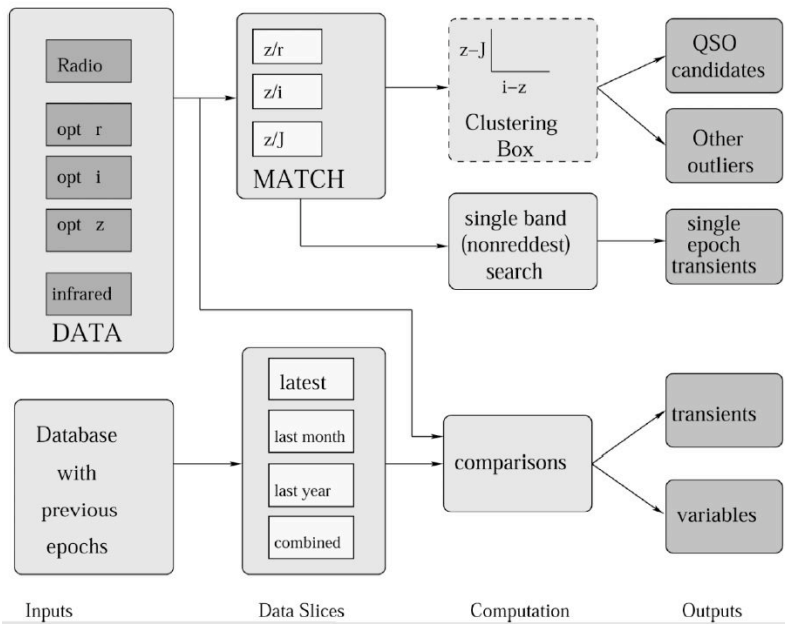

Figure 7. A schematic example of another GRIST application, for a search for transients and color-space outliers in the PQ survey. This is a concept of a next generation sky survey data analysis pipeline.

The GRIST system is currently in the prototype and testing stage. While this system is intended primarily for astronomical research, the underlying philosophy is generally applicable to other scientific applications.

Many of the key data mining and analysis tasks in design and implementation of such systems are in the broad arena of pattern recognition, similar to the examples described above. They include automated classification and clustering problems such as those described above, outlier or anomaly searches, correlation searches (effectively clustering analysis with a reduction of statistical dimensionality), and effective visualization of high-dimensionality and/or highly complex data spaces and constructs. We have an urgent need for a new generation of highly scalable algorithms for such tasks, which will become the centerpieces of the new generation of computational data systems.

## 5. Concluding Comments

All sciences today are being profoundly transformed by the advances in information technology. Yet, it is fair to say that we are not yet making a full use of the remarkable richness of modern massive and complex data sets. As they grow exponentially, so does the challenge of knowledge extraction from them.

These problems are not merely technical, but are deeply intellectual [3,10]. We are seeing development of a new scientific methodology suitable for the exploration and exploitation of massive and complex data sets and simulations. In a way, the situation is similar to the introduction of statistics in science in the $18^{th}$ century, and the development of modern experimental method and design ever since. We build on the foundations of the past, but we need some new tools and techniques. Applied computer science is increasingly playing the role which mathematics played since Newton. All science is now becoming "eScience" or "Cyberscience" – useful labels in the transition period, but soon to be quaint, as Petabytes and Petaflops become the norm.

The various computational and data challenges and opportunities we see in astronomy are fairly universal and applicable to many other fields. This is an area in which there is much room and need for collaborations between domain scientists such as astronomers, and computer scientists, statisticians, and IT experts.

Pattern recognition is a useful intellectual metaphor to describe the scientific discovery process as it happens in the mind of a scientist. As the empirical basis of science, in the form of massive and complex data sets and constructs, starts to exceed the intuitive and perhaps even cognitive capacity of the human mind, we could do well to develop some machine assistance in this process.

## 6. Acknowledgments

We are grateful to C. Baltay, D. Rabinowitz and other members of the PQ Survey team, M. Stalzer and the support staff at Caltech CACR, J. Jacob at JPL, and numerous collaborators and colleagues involved in various VO-related projects. This work was supported in part by the U.S. NSF grants AST-0407448, AST-0326524, CNS-0540369, AST-0122449. SGD also thanks the Ajax Foundation for support, and acknowledges the hospitality of EPFL and Geneva Observatory, where some of this paper was completed.## 7. References

[1] A. Szalay, & J. Gray 2001, "The World-Wide Telescope", *Science*, **293**, 2037-2040.

[2] R. Brunner, S.G. Djorgovski, T.Prince, & A. Szalay, "Massive Data Sets in Astronomy," in *Handbook of Massive Data Sets*, eds. J. Abello et al., Dordrecht: Kluwer Academic Publ., 2002, pp. 931-979. http://www.arXiv.org/abs/astro-ph/0106481

[3] S.G. Djorgovski, "Virtual Astronomy, Information Technology, and the New Scientific Methodology", in *IEEE Proc. of CAMP05: Computer Architectures for Machine Perception*, eds. V. Di Gesu & D. Tegolo, 2005, p. 125. http://www.arXiv.org/abs/astro-ph/0504651

[4] U.S. NVO White Paper, "Toward a National Virtual Observatory: Science Goals, Technical Challenges, and Implementation Plan", *A.S.P. Conf. Ser*. vol. **225**, p. 353, (2001). http://www.arXiv.org/abs/astro-ph/0108115

[5] R. Brunner, S.G. Djorgovski, & A. Szalay (editors), *Virtual Observatories of the Future*, A.S.P. Conf. Ser. Vol. **225**, San Francisco: Astron. Soc. of the Pacific, 2001.

[6] The National Virtual Observatory Science Definition Team report, available at http://nvosdt.org

[7] S.G. Djorgovski & R. Williams 2005, "Virtual Observatory: From Concept to Implementation", in: *From Clark Lake to the Long Wavelength Array*, eds. N. Kassim et al., A.S.P. Conf. Ser. Vol. **345**, p. 517. http://www.arXiv.org/abs/astro-ph/0504006

[8] The U.S. National Virtual Observatory (NVO) project website: http://us-vo.org

[9] The International Virtual Observatory Alliance (IVOA) website: http://ivoa.net

[10] D. Atkins, et. al., *Revolutionizing Science and Engineering through Cyber-Infrastructure*, Report of the NSF Blue-Ribbon Advisory Panel on Cyberinfrastructure, 2003, http://www.nsf.gov/od/oci/reports/toc.jsp

[11] The Palomar-Quest Digital Sky Survey website: http://www.astro.caltech.edu/pq/

[12] S. C. Odewahn, et al. 1992, "Automated star/galaxy discrimination with neural networks", *Astron.J.*, **103,** 318.

[13] N. Weir, U. Fayyad, & S.G. Djorgovski 1995, "Automated Star/Galaxy Classification for Digitized POSS-II", *Astron. J.,* **109**, 2401.

[14] S. C. Odewahn, et al. 2005, "The Digitized Second Palomar Observatory Sky Survey (DPOSS). III. Star-Galaxy Separation", *Astron. J.,* **128**, 3092.862